\documentclass[aps,twocolumn,,superscriptaddress,showpacs,showkeys,amsmath,amssymb,floatfix]{revtex4}
\usepackage{graphicx}
\usepackage{epsfig}
\usepackage{dcolumn}
\usepackage{bm}
\usepackage{amssymb}
\usepackage{dsfont}
\usepackage{amsmath}
\usepackage{subfigure}
\newcommand{\be}{\begin{equation}}
\newcommand{\ee}{\end{equation}}
\newcommand{\bea}{\begin{eqnarray}}
\newcommand{\eea}{\end{eqnarray}}
\newcommand{\hn}{\hat n}

\newcommand{\tC}{{\tilde C}}

\newcommand{\pro}{\partial}
\newcommand{\der}{\partial}

\newcommand{\ba}{\begin{array}}
\newcommand{\ea}{\end{array}}

\newcommand{\nn}{\nonumber}

\newcommand{\uast}{\stackrel{\ast}{u}}

\begin{document}
\title{Monopoles without magnetic charges: Finite energy monopole-antimonopole
configurations in $CP^1$ model and restricted QCD}
\bigskip
\author{L. P. Zou}
\affiliation{Research Center for Hadron and CSR Physics,
Lanzhou University and Institute of Modern Physics of CAS, Lanzhou 730000, China}
\affiliation{Institute of Modern Physics, Chinese Academy of Sciences,
Lanzhou 730000, China}
\author{D. G. Pak}
\affiliation{Institute of Modern Physics, Chinese Academy of Sciences,
Lanzhou 730000, China}
\affiliation{Lab. of Few Nucleon Systems,
Institute for Nuclear Physics, Ulugbek, 100214, Uzbekistan}
\author{P. M. Zhang}
\affiliation{Research Center for Hadron and CSR Physics,
Lanzhou University and Institute of Modern Physics of CAS, Lanzhou 730000, China}
\affiliation{Institute of Modern Physics, Chinese Academy of Sciences,
Lanzhou 730000, China}
\affiliation{State Key Laboratory of Theoretical Physics, Institute of Theoretical Physics, 
Chinese Academy of Sciences, Beijing 100190, China}
\begin{abstract}
We propose a new type of regular monopole-like field configuration
in quantum chromodynamics (QCD) and $CP^1$ model. The monopole configuration
can be treated as a monopole-antimonopole pair without localized magnetic charges.
An exact numeric solution for a simple monopole-antimonopole
solution has been obtained in $CP^1$ model with an appropriate
potential term. We suppose that similar monopole solutions may exist
in effective theories of QCD and in the electroweak standard model.
\end{abstract}
\vspace{0.3cm}
\maketitle

\section{Introduction}

One of most attractive mechanisms of quark confinement in
quantum chromodynamics is based on Meissner effect in
dual color superconductor \cite{nambu,mandelstam,polyakov77}.
This assumes generation of monopole vacuum condensate due to quantum dynamics of gluons.
The existence of the monopole condensation represents a
puzzle in QCD which is intimately related with
the problem of origin of the mass gap in QCD.
Due to correspondence principle between quantum and classical
descriptions one would expect that QCD admits classical
monopole solutions. However, since invention of singular
Dirac \cite{dirac} and Wu-Yang monopoles \cite{wuyang}
up to present moment we don't know any finite energy
monopole solutions in realistic physical
theories of fundamental interactions.
All known finite energy monopole solutions
(like 't Hooft-Polyakov one \cite{thooft,polyakov74})
require either introducing new particles or
essential extension of the theories of fundamental forces.

The problem of existence of finite energy monopoles in QCD
becomes more critical due to the following: the
magnetic potential in $SU(2)$ QCD is given by dual Abelian gauge potential
which is defined in terms of $CP^1$ field $\hat n$ \cite{choprd80,duan}.
The magnetic potential is divergenceless, this implies that any monopole field configuration
defined in terms of $\hat n$ with non-vanishing magnetic charge unavoidably
contains singularities on some subset of three dimensional space.
One possibility to overcome this problem
is to introduce additional fields as it occurs in the case of composite
't Hooft-Polyakov monopole.
Another way to avoid such singularities is
to consider monopole configurations with vanishing total magnetic
charge. This approach is based on the fact that monopole charge
in pure QCD does not represent gauge invariant quantity, so,
since the color symmetry is unbroken only
monopole configurations with a total zero magnetic charge can
serve as a classical analog to gauge invariant monopole vacuum condensate.
Moreover, it has been found that monopole-antimonopole string (or knot) pair
can form a stable vacuum in QCD \cite{pakplb06}.

In this Letter we study the problem of existence
of finite energy monopoles in QCD considering a subsector
of standard QCD which is defined formally by the $CP^1$ Lagrangian.
We propose a new type of finite energy monopole field configuration
which can be treated as a monopole-antimonopole pair.
An essential feature of such monopole configuration is that it
does not have localized magnetic charge anywhere in contrast to
known monopole-antimonopole solutions in Yang-Mills-Higgs theory.
We consider first singular monopole-antimonopole configuration
which represents a limiting case of the system of
point-like monopole and antimonopole approaching each other
at zero distance. It is not much surprising that such a singular
non-trivial configuration takes place.
Unexpected result is that there exists a finite energy
monopole-antimonopole configuration which is regular everywhere
and which minimizes the energy functional of restricted QCD.

The paper is organized as follows.
In Section II we consider singular monopoles in $CP^1$ model.
In Section III we study the structure of regular finite energy
monopole-antimonopole configurations. To demonstrate that such a configuration
can be realized as a solution we obtain exact numeric solution
for a simple monopole-antimonopole field in a simple
$CP^1$ in Section IV.
\section{Singular monopole-antimonopole pair in $CP^1$ model}
Let us consider a simple $CP^1$ model defined by the Lagrangian
\bea
{\cal L}_0&=&-\dfrac{1}{4}H_{\mu\nu}^2,   \label{lagrcp1}
\eea
where the magnetic field $H_{\mu\nu}$ is expressed in terms of the $CP^1$ field $\hat n$
as follows
\bea
H_{\mu\nu}&=&\epsilon^{abc} \hat n^a \der_\mu \hat n^b \der_\nu \hat n^c.
\eea
The Lagrangian ${\cal L}_0$ describes a subsector
of the standard QCD with a restricted
$SU(2)$ gauge potential given by
\bea
\hat A_\mu=-\dfrac{1}{g} \hat n \times \pro_\mu \hat n.
\eea
The expression for the gauge potential
can be treated either as a reduction of
the standard QCD to the restricted QCD \cite{choprd80,duan},
or as a special ansatz in Faddeev-Niemi approach to
formulation of the effective theory of QCD in the
infra-red limit \cite{FNprl99,shabanov}. In first case the field $\hat n$ satisfies
the equations of motion of the standard QCD, whereas in the second case
the equations of motion for $\hat n$ are determined exactly
by the $CP^1$ Lagrangian ${\cal L}_0$.


It is convenient to express the $CP^1$ field through the
complex field $u(x)$ using stereographic projection
\bea
\hat n &=& \dfrac{1}{1+u \uast}
 \left (\ba{c}
  u+\uast\\
  -i (u-\uast)\\
 u \uast-1\\
            \ea
           \right ). \label{nstereo}
\eea
With this the magnetic field $H_{\mu\nu}$ can be written
explicitly in terms of $u$
\bea
H_{\mu\nu}&=& \dfrac{-2 i}{(1+|u|^2)^2}
 (\der_\mu u \, \der_\nu \uast-\der_\nu u \,\der_\mu \uast).
 \eea
The magnetic field $H_{\mu\nu}$ determines a corresponding
closed differential 2-form $H=dx^\mu \wedge dx^\nu H_{\mu\nu}$
which implies local existence of a
dual magnetic potential $\tC_\mu$
\bea
&& H_{\mu\nu} = \pro_\mu \tC_{\nu }-\pro_\nu \tC_{\mu }. \label{dualpot}
\eea
All topologically non-equivalent configurations of
the $CP^1$ field $\hat n$ are classified by homotopy groups
$\pi_{2,3}(CP^1)$. Consequently one can classify possible topological
fields $\hat n$ by Hopf, $Q_H$, and monopole, $g_m$, charges
\bea
Q_H&=& \dfrac{1}{32 \pi^2}\int d^3x \epsilon^{ijk} \tC_i H_{jk} , \nn \\
g_m&=&\dfrac{1}{V(S)}\int_{S^2} H_{ij} \cdot d \sigma^{ij},
\eea
where $V(S)$ is a volume of the sphere $S^2$.

The magnetic vector field $\vec H^i=\dfrac{1}{2}\epsilon^{ijk} H_{jk}$
has a vanishing divergence. This implies that a regular finite energy
monopole configuration with a non-zero magnetic charge does not exist
in a simple $CP^1$ model and in the restricted QCD
unless one introduces additional fields to make a composite monopole.
Monopole-like field configurations with a total vanishing magnetic charge
can be in principle regular everywhere since in that case there is no
any topological obstructions.
Monopole-antimonopole pairs made of Dirac or Wu-Yang monopoles
\cite{nambu77} still possess singularities in the centers of the point-like monopoles.
Besides, such a pair does not represent a static bound state, since
obviously monopole and antimonopole will annihilate and disappear.
However, in theories with non-linear self-interaction one may expect existence
of a non-trivial monopole configuration in the limit when
monopole and antimonopole approach each other at zero distance.
Indeed, we will show that such a singular field configuration exists in $CP^1$ model.

 Let us define the following axially symmetric ansatz
\bea
u(r,\theta,\varphi)&=& e^{i m \varphi} \big (\cot(\dfrac{n\theta}{2}) f(r,\theta)+
i \csc (\dfrac{n\theta}{2})Q(r,\theta)\big), \nn \\
&& \label{ansmm}
\eea
where the integer numbers $(m,~n)$ are winding numbers corresponding to the spherical
angles $(\varphi,~\theta)$. A simple setting $m=n=1, ~f=1,~  Q=0$ reproduces Wu-Yang monopole
solution with a unit magnetic charge, $g_m=1$. Another interesting case with
winding numbers $m=1,~ n=2$ and given functions
\bea
f(r,\theta)=1,~~~~~~ Q(r,\theta)=\dfrac{a^2-r^2}{2 a r}
\eea
leads to exact knot solitons with Hopf charge $Q_H=1$
in a special integrable $CP^1$ model
\cite{nicole, AFZ} and in a generalized Skyrme-Faddeev model \cite{zzpprd13}.
We will consider the case $m=1,~n=2$ with an appropriate choice of functions
$(f,~Q)$ leading to zero total magnetic charge, the value of the Hopf charge
depends on imposed boundary.
It turns out that the ansatz (\ref{ansmm}) provides a rich structure of possible
monopole like configurations admitting non-trivial twisted magnetic fluxes.

Let us first consider general properties of magnetic field configurations
with the ansatz (\ref{ansmm}).
For numeric purpose it is convenient to change the variables
\bea
f(r,\theta)&=&2-\dfrac{1}{F(r,\theta)}, \nn \\
Q(r,\theta)&=& \dfrac{1}{G(r,\theta)}-1.
\eea
Asymptotic properties of the magnetic field is determined by values
of the functions $(f,Q)$ and their first derivatives
at infinity $r \rightarrow \infty$. To find proper boundary conditions
near origin and near infinity we consider simple case of radial dependent functions
$f(r),~Q(r)$. With this one can write the magnetic field components as follows
\bea
H_{r\theta}&=& -\dfrac{4 \sin \theta (f Q'-\cos^2 \theta f'Q)}{(\cos^2\theta f^2+Q^2+\sin^2\theta)^2},\nn \\
H_{r\varphi}&=&\dfrac{4 \sin^2 \theta (\cos^2 \theta ff'+Q')}{(\cos^2\theta f^2+Q^2+\sin^2\theta)^2},\nn \\
H_{\theta\varphi}&=& -\dfrac{2 (f^2+Q^2) \sin (2\theta)}{(\cos^2\theta f^2+Q^2+\sin^2\theta)^2}.
\eea
In asymptotic region near infinity, $r\rightarrow \infty$, one has
\bea
H_{r\theta}&\rightarrow& 0, \nn \\
H_{r\varphi}&\rightarrow& 0, \nn \\
H_{\theta\varphi}&\rightarrow&-\dfrac{2 (f_\infty^2+Q_\infty^2)
       \sin (2\theta)}{(\cos^2\theta f_\infty^2+Q_\infty^2+\sin^2\theta)^2},
\eea
where $f(r=\infty)=f_\infty,~ Q(r=\infty)=Q_\infty$.

Let us choose the following following boundary conditions
written for the functions $F,~G$:
\bea
F(0)=F(\infty)=\dfrac{1}{2}, ~~~~~~G(0)=G(\infty)=\dfrac{1}{2}.
\eea
One can calculate magnetic fluxes through the upper and lower semi-spheres
$H^2_\pm$ of the sphere $S^2$ od radius $R$ centered at the origin $r=0$.
In the asymptotic limit $R\rightarrow \infty$ the magnetic fluxes are given by
\bea
\Phi_+ &=&\int_0^{2 \pi} d\varphi \int_0^{\pi/2} d\theta H_{\theta\varphi}=-2 \pi, \nn \\
\Phi_- &=&\int_0^{2 \pi}d\varphi\int_{\pi/2}^\pi d\theta H_{\theta\phi}=+2 \pi.
\eea
The Hopf charge for the given configuration is zero.
The magnetic fluxes are multiples of the minimal magnetic flux quantum $2 \pi$
for arbitrary non-zero value of $F(0)$.
The field configuration
looks like monopole (antimonopole) for
observers standing at far distance in lower (upper) half-space.
Notice, the configuration has a singularity at the point $r=0$.
One can consider the magnetic flux
through the surface composed from an upper semi-sphere
of small radius $\rho_0$ centered at the origin $r=0$,
$(\theta \in [0,\pi/2], \varphi \in [0, 2 \pi])$,
 and a disc $D^2: \{r\leq \rho_0\}$ in the
$(X,Y)$ plane
\bea
\Phi_{0+}&=&\int_0^{2 \pi} d\varphi \int_0^{\pi/2} d\theta H_{\theta\varphi}(\rho_0,\theta) \nn \\
         &+&\int_0^{2 \pi} d\varphi
\int_0^{\rho_0}dr H_{r\varphi}(r,\pi/2).
\eea
One can check that the magnetic flux $\Phi_{0+}$ converges to a value
$-2 \pi$ in the limit $\rho_0 \rightarrow 0$.
Similarly, the
magnetic flux $\Phi_{0-}$ through the surface made of a lower semi-sphere of radius
$\rho_0$, $(\theta \in [\pi/2,\pi],\varphi \in [0, 2 \pi])$, and a disc $D^2$,
converges to a value $+2 \pi$ when $\rho_0 \rightarrow 0$.
Such a behavior corresponds to the point-like monopole and antimonopole
placed in one point, $r=0$.
We will treat such a configuration
as a monopole-antimonopole pair. Notice, that our monopole configuration
is essentially non-Abelian, and it is different from the
monopole-antimonopole pair of Dirac monopoles forming a magnetic dipole
due to superposition rule available in linear field theory.

The semi-spheres $H^2_{\pm}$ have a common boundary $S^1$ as a circle
in the $(X,Y)$ plane, and due to axial symmetry the vector field $\hat n$
is a constant vector along the boundary.
One can make compactification of the semi-spheres $H_{\pm}$
to spheres $S_{\pm}$ by identifying all points of the boundary
to one point. With this it becomes clear that magnetic flux quantization originates
from the non-trivial mappings $\pi_2(S^2)$  corresponding to monopole
and antimonopole.
In asymptotic region $r\rightarrow \infty$ the $CP^1$ field
can be written as follows
\bea
\hn_1&=&\sin\theta(\cos\varphi \cos\theta-\sin\varphi), \nn \\
\hn_2&=&-\sin\theta(\cos\varphi+\cos\theta\sin\varphi), \nn \\
\hn_3&=&\cos^2\theta.
\eea
This implies that $\hat n$ represents
twisted one-to-one mapping $S^2\rightarrow S^2$.

The monopole-antimonopole configuration
resembles a finite energy solution for monopole-antimonopole in Yang-Mills-Higgs
theory \cite{KKprd} where monopole and antimonopole are localized
at zeros of the Higgs field on the $Z$-axis.
One should stress that our monopole-antimonopole differs from the
monopole-antimonopole in Yang-Mills-Higgs theory which represents a composite
monopole while our monopole represents a pure monopole system without any additional
scalar fields.
The construction of the singular monopole-antimonopole configuration
gives a hint that a finite energy monopole-antimonopole configuration might exist
even in a simple $CP^1$ model. The key point is that the fields $f$ and $Q$ can regularize
the singularity through the "dressing" effect in a similar manner as the
Higgs field regularizes the singularity at the origin in the
't Hooft-Polyakov monopole solution.
Surprisingly, we have found that such a regular monopole configuration is allowed
in the $CP^1$ model and consequently in QCD.
\section{Regular finite energy monopole-antimonopole configuration}

For simplicity we consider radial dependent functions
$f(r),~Q(r)$ ($(F(r),~ G(r)$).
Let us impose the following boundary conditions
\bea
&&F(0)=0,~~~~~F(\infty)=1,\nn\\
&&G(0)=0,~~~~~G(\infty)=1.
\eea
The $CP^1$ field $\hat n$ behaves like a Higgs field taking
constant vector value at the origin
\bea
\hn(r=0)=(0,0,1).
\eea
In asymptotic region $r\rightarrow \infty$ the
vector field $\hat n$ can be written as follows
\bea
\hn=(\cos \varphi \sin(2 \theta),~-\sin \varphi \sin(2 \theta),~ \cos(2 \theta)),
\eea
i.e., the field $\hat n$ provides
double covering of the sphere $S^2\simeq CP^1/U(1)$.
One can calculate
the magnetic fluxes through the upper and lower semi-spheres
$H_\pm^2$ of the sphere $S^2$ of infinite radius
\bea
\Phi_+ &=&\int_0^{2 \pi} d\varphi \int_0^{\pi/2} d\theta H_{\theta\varphi}=-4 \pi, \nn \\
\Phi_- &=&\int_0^{2 \pi} d\varphi \int_{\pi/2}^\pi d\theta H_{\theta\varphi}=+4 \pi. \label{fluxes2a}
\eea
One has non-zero magnetic flux around the $Z$-axis created by the magnetic field
$H_{r\theta}$  which implies a helical structure of the monopole
configuration
\bea
\Phi_{\varphi}=\int_0^\infty dr \int_0^\pi d\theta H_{r\theta}=2 \pi.
\eea
The twisted magnetic fluxes correspond to half-integer value of the Hopf charge,
$Q_H=\frac{1}{2}$.

In further we will study possible solutions with properties
of such monopole-antimonopole configurations in $CP^1$ model and electroweak theory.
Due to this it is important to verify strictly the regular structure
of the monopole configuration. To get a qualitative description
we use simple radial trial functions
\bea
F(r)&=&1+e^{-\frac{r}{2}}(-1+r^2), \nn \\
G(r)&=&1+e^{-\frac{r}{2}}(-1+r^2).
\eea
With this one can study the detailed structure of the magnetic field
for a given monopole-antimonopole configuration.
From the density contour plot of the magnetic vector field
projected onto the planes $(X,Z)$ and $(X,Y)$, Figs. 1,~2 respectively,
one can see that at far distance above (below) the plane $(X,Y)$ the magnetic flux corresponds to
negatively (positively) charged monopole
with the magnetic field lines twisted around the $Z$-axis.
\begin{figure}[htp]
\centering
\includegraphics[width=50mm,height=50mm]{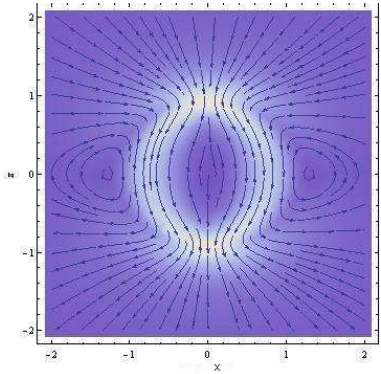}
\caption[CP1plotstrialZX]{Density contour plot of the magnetic vector field in the plane $y=0$.
}\label{fig:CP1trialZX}
\end{figure}
In Fig. 2.a-d the one can retrieve a non-trivial helical structure
of the magnetic field.
The magnetic vector field lines starting at far distance $z>>0$
approach the central area and wind around the $Z$-axis,
Fig. 2a.
Passing the plane $z=0.93$ the magnetic field lines are coming untwisted
and moving away from the $Z$-axis, Fig. 2b,c. In the interior
area $0\leq z\leq 0.93$  the magnetic fields approach the plane
$z=0$ with a quite complicate helical structure which shows opposite
winding directions above and below the plane $z=0$, Fig. 2d. In lower half-space
the magnetic field has a similar behavior due to the reflection symmetry $z\rightarrow -z$.
\begin{figure}[htp]
\centering
\subfigure[~$z=1.0$ ]{\includegraphics[width=40mm,height=40mm]{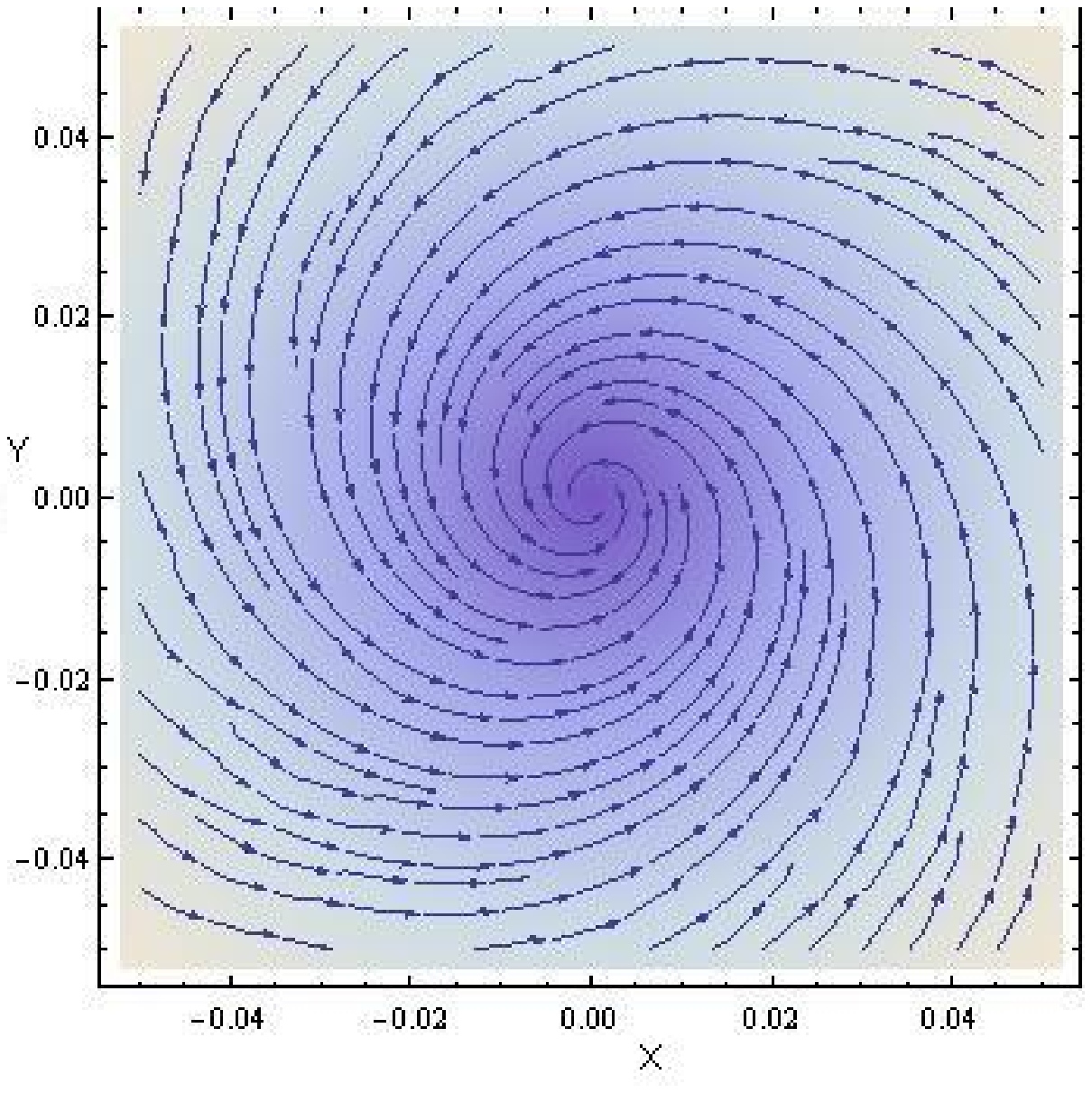}}
\hfill
\subfigure[~$z=0.93$ ]{\includegraphics[width=40mm,height=40mm]{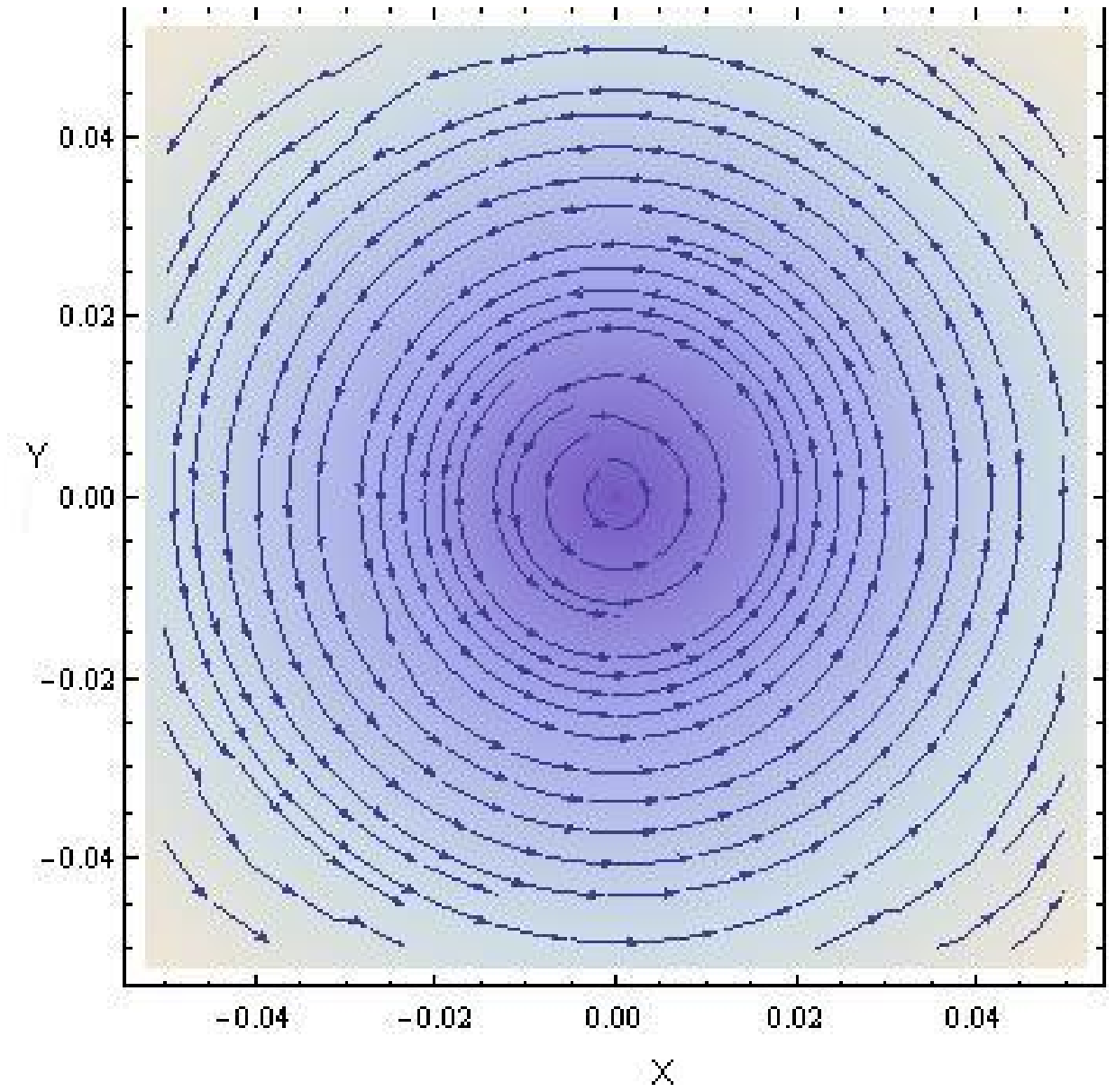}}\\
\subfigure[~$z=0.7$ ]{\includegraphics[width=40mm,height=40mm]{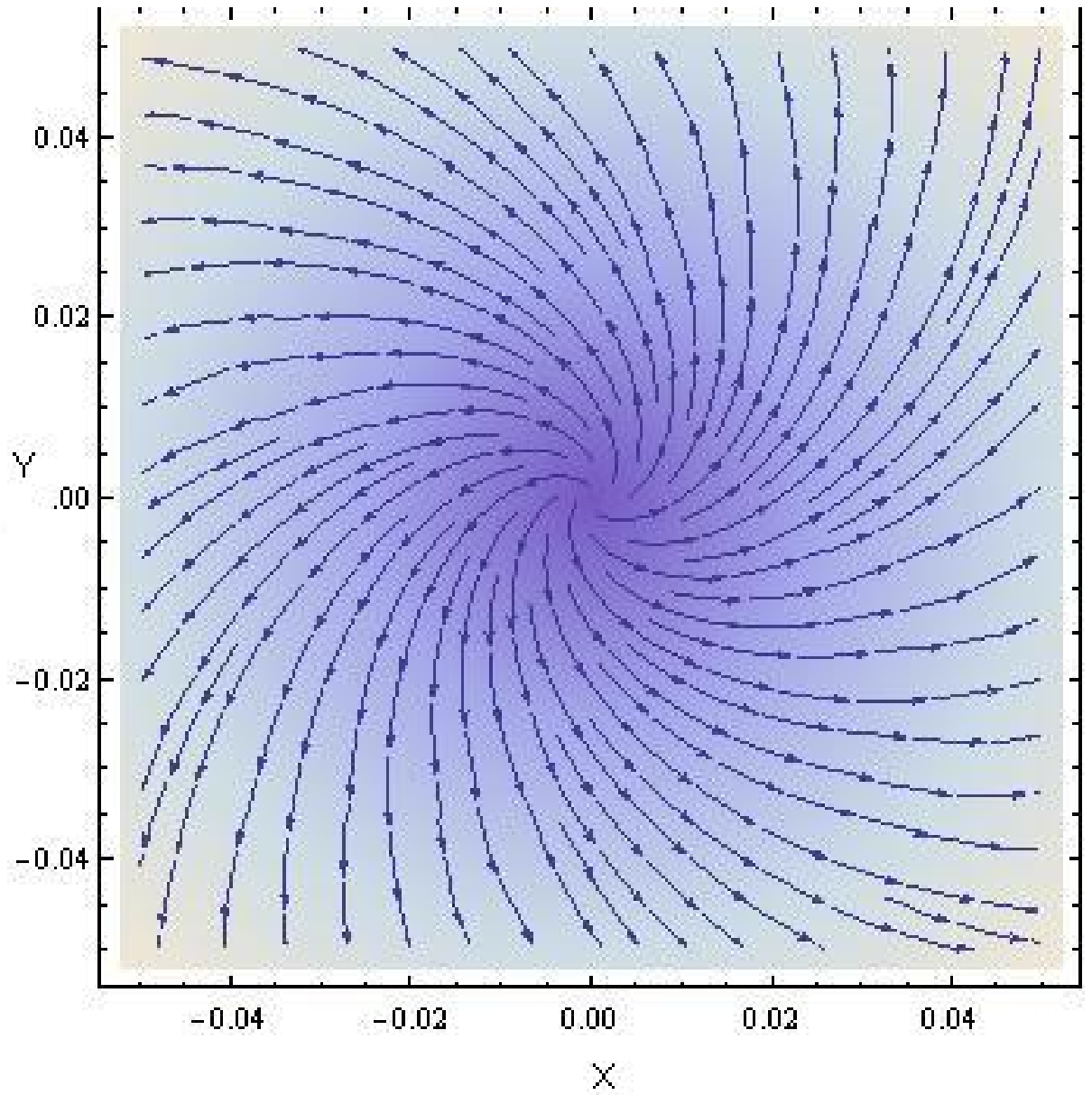}}
\hfill
\subfigure[~$z=+0.01$ ]{\includegraphics[width=40mm,height=40mm]{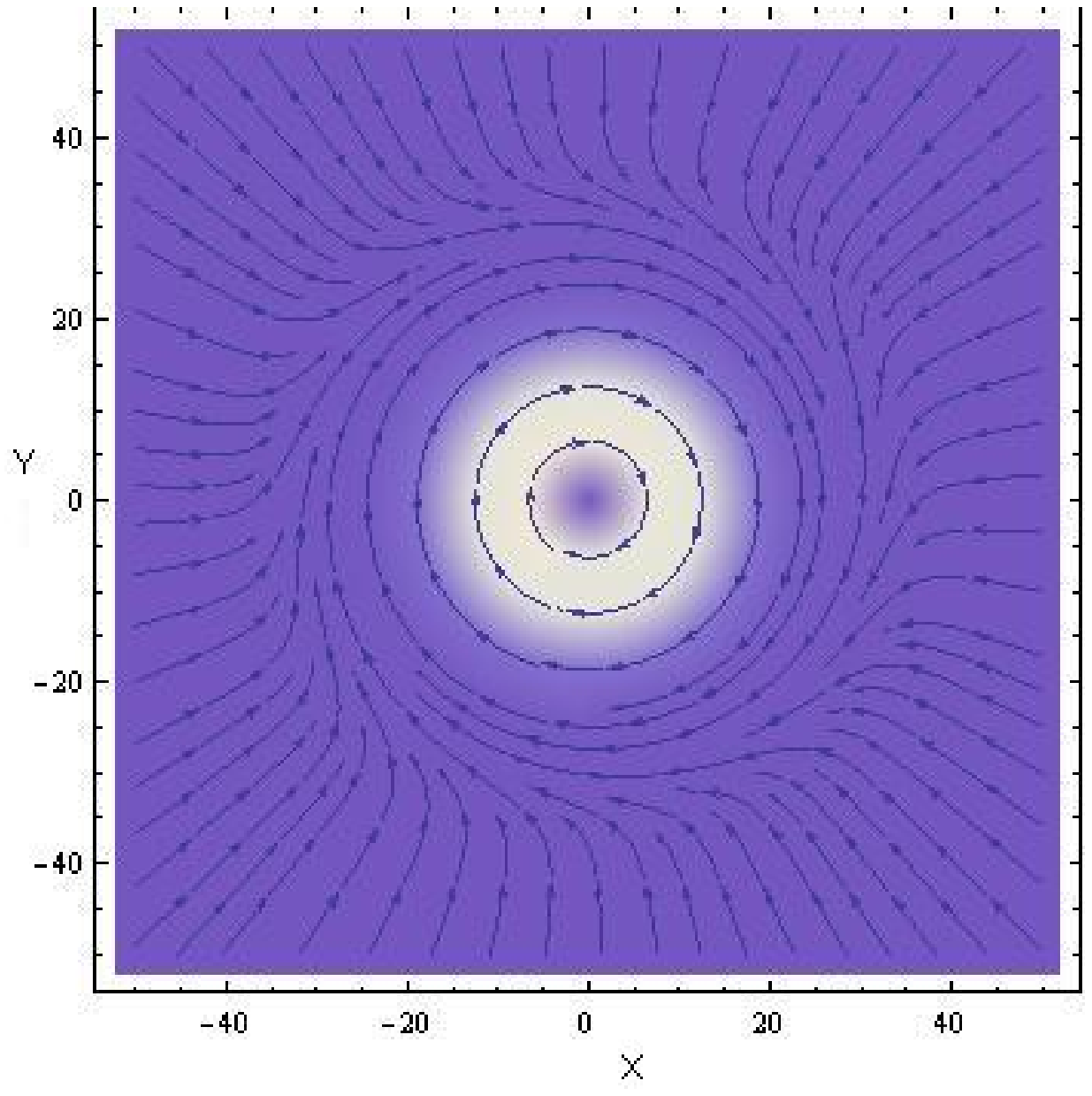}}
\caption[CP1plotstrialXY]{Density contour plots of the magnetic field
in the planes: (a) $z=1.0$; (b) $z=0.93$; (c) $z=0.7$; (d) $z=+0.01$.
}\label{fig:CP1trialXY}
\end{figure}
The vector stream plot for the magnetic field, Fig. 3, shows the general regular structure
and the presence of two local maximums of density of the magnetic field located
near the $Z$-axis in the planes $z=\pm 0.93$.
\begin{figure}[htp]
\centering
\includegraphics[width=55mm,height=55mm]{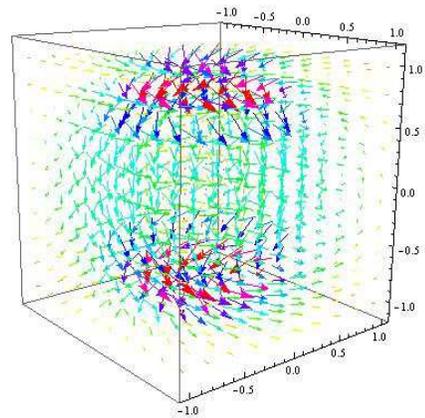}
\caption[CP1plotstrial3D]{Three-dimensional vector stream plot for the magnetic field
in Cartesian coordinates.
}\label{fig:CP1trial3D1}
\end{figure}

To make sure that magnetic field has a regular structure everywhere
we make plots for the vector lines which start from four symmetric points
in the upper half-space, Fig. 4.
Conditionally one can select two types of magnetic vector field lines.
The magnetic field lines of the first type are localized along the $Z$-axis,
Fig. 4a,b, and the magnetic field lines of the second type spread
to infinity along $x$ and $y$ directions.
\begin{figure}[htp]
\centering
\subfigure[~]{\includegraphics[width=40mm,height=40mm]{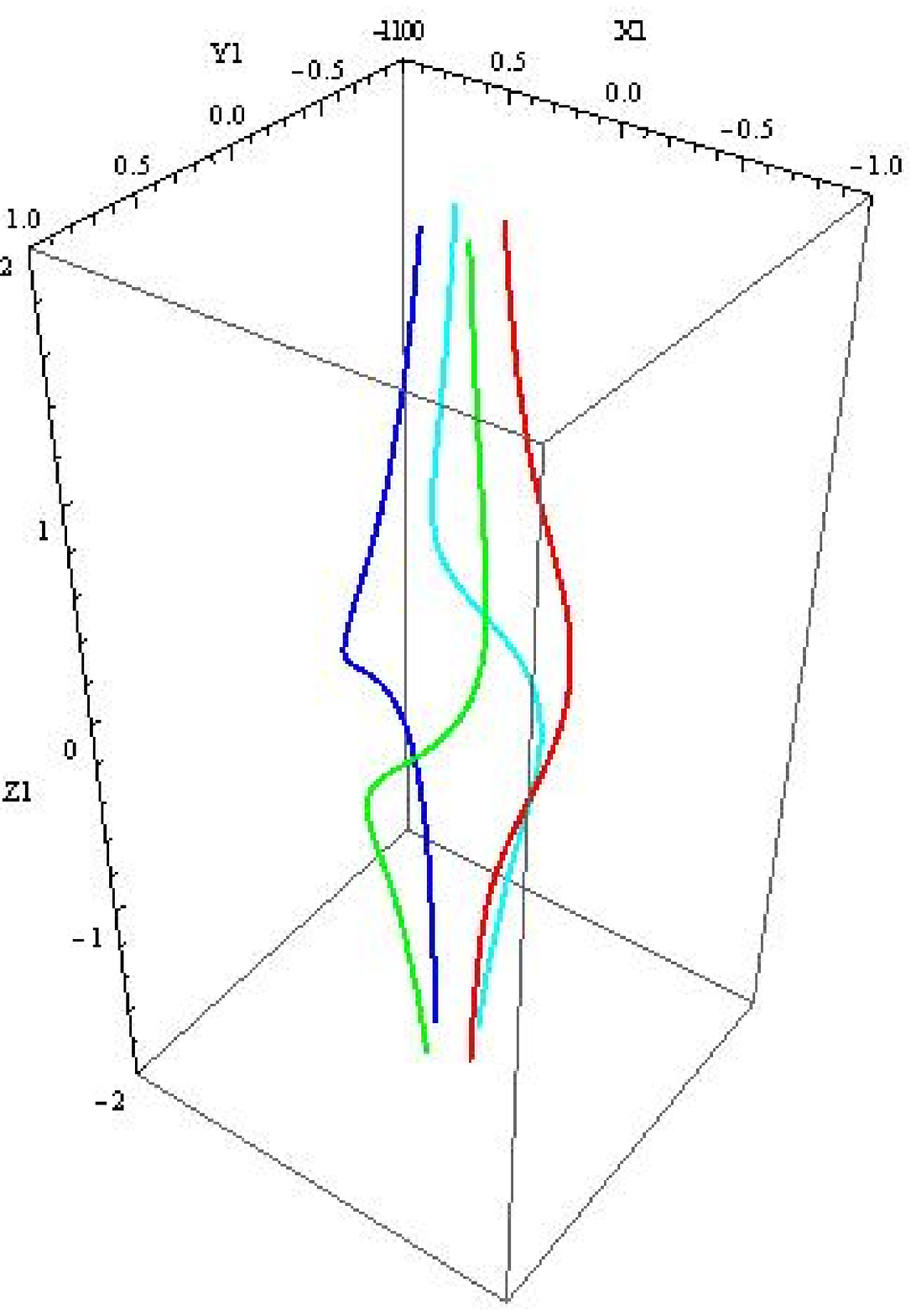}}
\hfill
\subfigure[~]{\includegraphics[width=42mm,height=44mm]{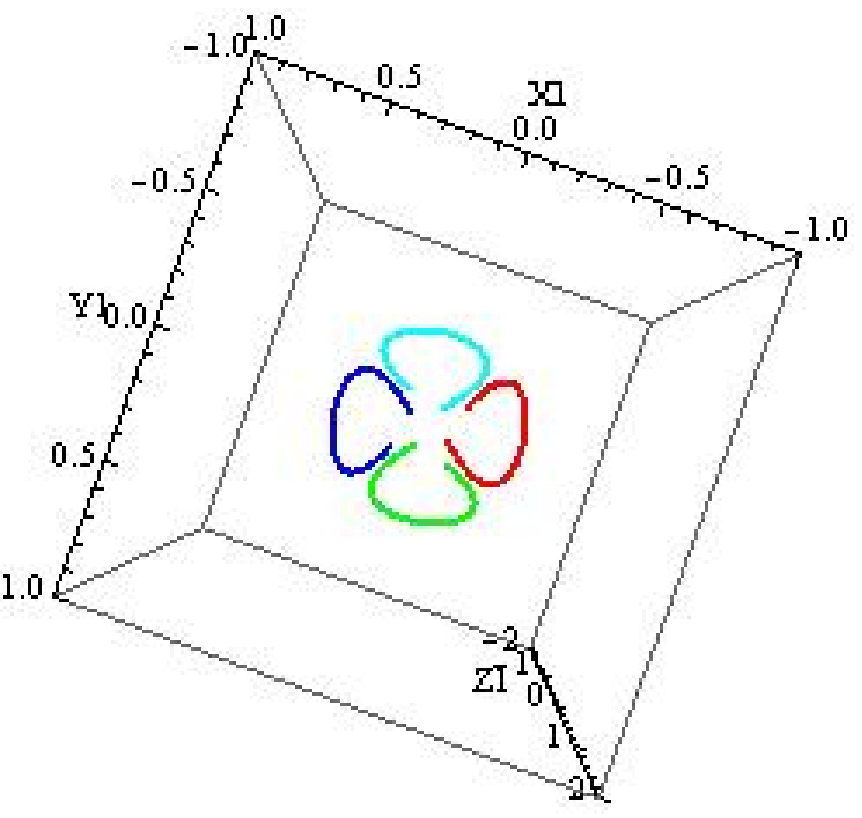}}\\
\subfigure[~]{\includegraphics[width=40mm,height=40mm]{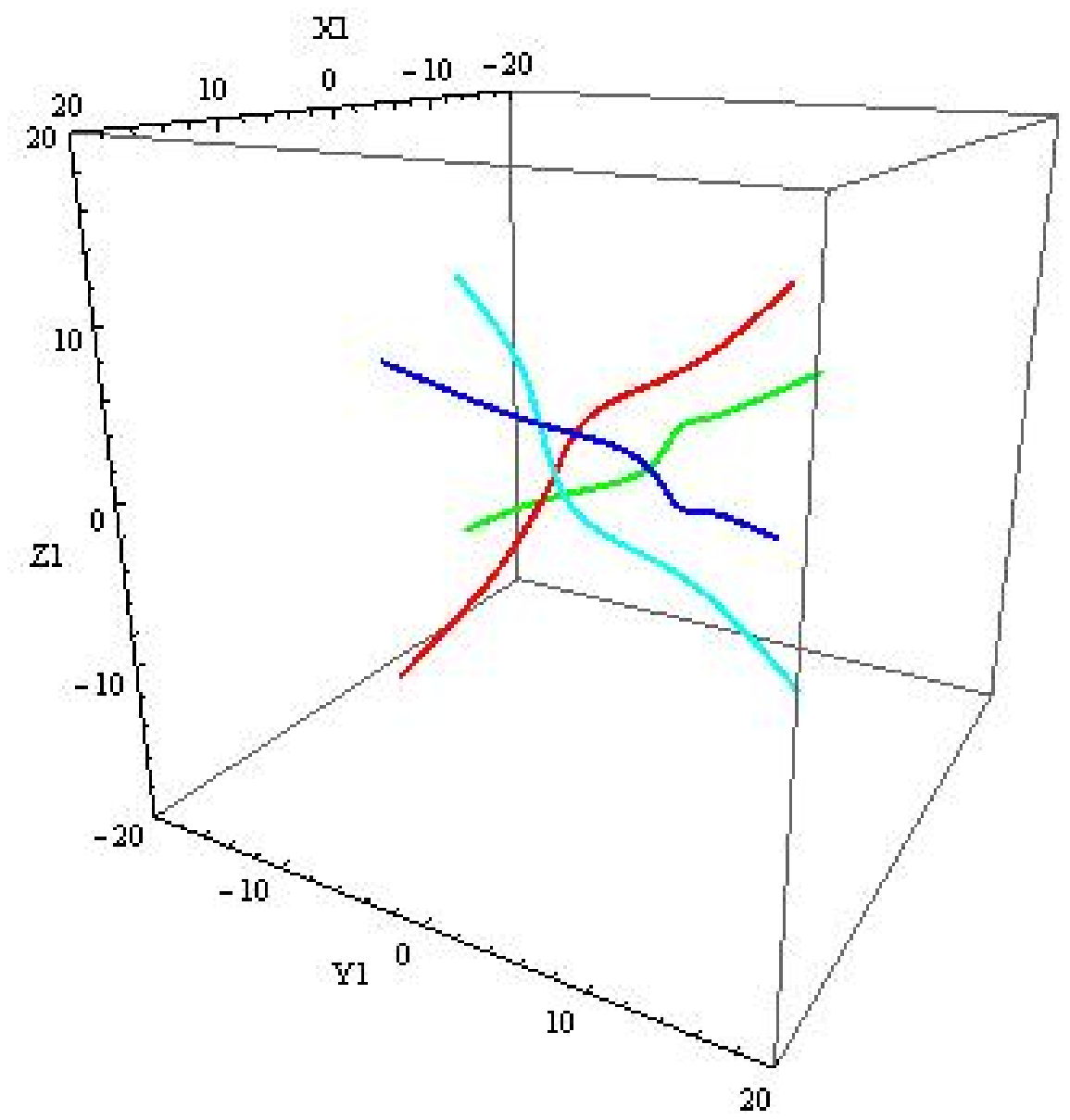}}
\hfill
\subfigure[~]{\includegraphics[width=40mm,height=40mm]{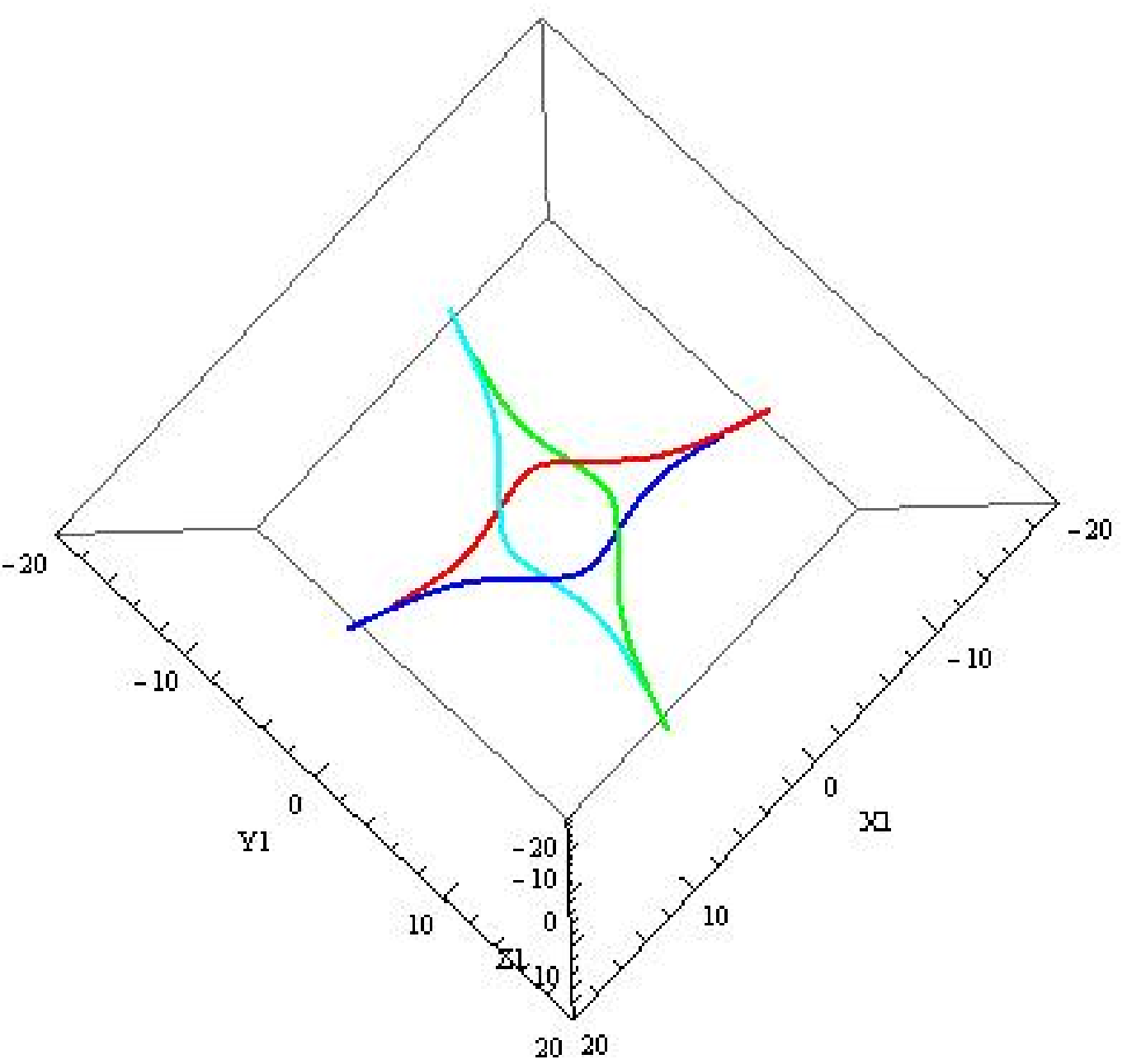}}
\caption[CP1plotstrialXY]{(a) Vector lines for the magnetic field $H_{mn}$
starting from four symmetric points near $Z$-axis; the same lines with the
point of view from above; (c) magnetic field vector lines starting from
four symmetric points at large distance from $Z$-axis; (d) the same lines
with the point of view from above.
}\label{fig:vectorlines}
\end{figure}
One can see from Fig. 5 that field configuration has two local energy density maximums
located in the planes $z=\pm 0.93$ along the circles $\theta \simeq 0.5,~\theta \simeq \pi-0.5$.
Maximal energy density forms two tori, so the configuration can be viewed
as a pair of monopole and antimonopole knots.
Notice, the given monopole-antimonopole configuration
has no localized magnetic charges anywhere contrary to the case of known
monopole-antimonopole solutions in Yang-Mills-Higgs theory \cite{KKprd},
i.e., the magnetic flux through any closed two-dimensional surface vanishes identically.
\begin{figure}[htp]
\centering
\includegraphics[width=65mm,height=55mm]{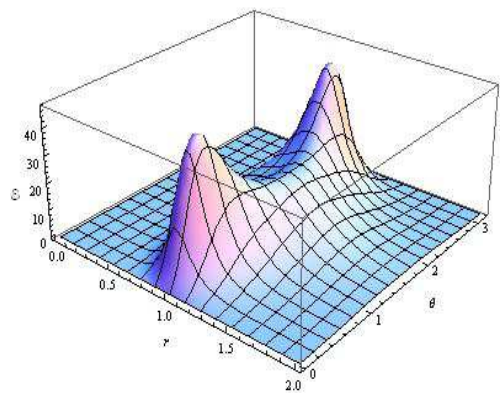}
\caption[CP1plotstrial3D]{Energy density plot for the magnetic field
in spherical coordinates $(r,\theta)$ (axially symmetric case).
}\label{fig:CP1trial3D2}
\end{figure}

In general, a detailed structure of the monopole-antimonopole configuration
can vary depending on dynamics determined by the equations of motion.
In the next section we will consider a more simple monopole-antimonopole
solution in $CP^1$ model without helical structure.

\begin{figure}[htp]
\centering
\includegraphics[width=65mm,height=55mm]{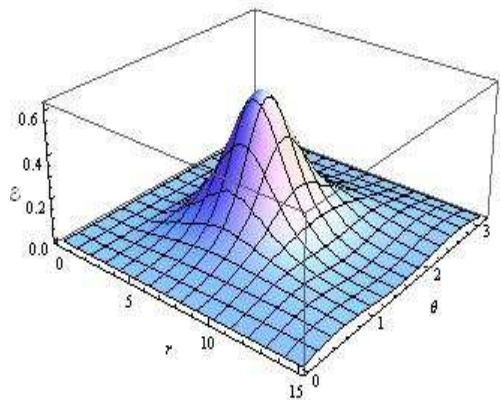}
\caption[CP1plotstrial3D]{Energy density plot for the magnetic field
configuration which minimizes the energy functional of restricted QCD.
}\label{fig:Endens5}
\end{figure}

Due to Derrick theorem \cite{derrick} the simple $CP^1$ model with the Lagrangian
(\ref{lagrcp1}) and the pure QCD do not admit a stable static solution. So
it is unlikely that monopole can be realized as a classical solution in standard
QCD. However, it is surprising that monopole-antimonopole configuration
considered above provide a minimum
of the energy functional for any given
total energy value. So that, one can not exclude completely the
possibility of existence of monopole-antimonopole solution in pure QCD.
To apply minimization procedure to the energy functional it is convenient
to pass to dimensionless variables $\tilde x^i$ by
rescaling $x^i \rightarrow d \tilde x^i$ with the length parameter $d$.
The trial variational functions $(F(r,\theta),~G(r,\theta))$
are chosen as Fourier series in $\sin k\theta, \cos l \theta$, ($k=1,3; l=2;4$),
with Laguerre polynomial radial coefficient functions $L_n(r)$, ($n=1,...5$).
In dimensionless variables the energy functional reads $(i,j=1,2,3)$
\bea
E&=&\dfrac{1}{4d}\int d^3 \tilde x \tilde H_{ij}^2.
\eea
We impose the following boundary conditions
\bea
F(0,\theta)&=&0,~~~~~G(0,\theta)=0, \nn\\
F(\infty,\theta)&=&1,~~~~~G(\infty,\theta)=1.
\eea
Minimizing procedure of the energy functional
gives the total energy value
\bea
E\simeq \dfrac{4.06}{d}.
\eea
The energy density plot, Fig. 6, shows that the monopole and antimonopole
merge together forming a toroidal structure. Since the Hopf charge is half-integer,
$Q_H=\frac{1}{2}$, the monopole-antimonopole configuration differs from
the topological knot and has non-vanishing magnetic fluxes through
the upper and lower semi-spheres $H_{\pm}^2$ of infinite radius, (\ref{fluxes2a}).
It is obvious that the configuration
can not represent a stable solution due to presence of the scale parameter $d$
which characterizes the effective size of the monopole configuration.
This is in close analogy with 't Hooft-Polyakov
monopole case where the energy of 't Hooft-Polyakov monopole in BPS limit
includes a scale parameter (averaged value of the Higgs field)
and its presence destabilizes the solution.

At far distance the given magnetic field configuration is similar
to a monopole-antimonopole bound state, and it does not possess
a localized magnetic charge inside any closed surface.
This type of solution can resolve a puzzle of existence of monopole
in QCD and in electro-weak standard theory, where, as it is well known, finite energy
non-composite monopole solution has not been found so far.
Due to possible importance of this monopole configuration
we will demonstrate the existence of such a solution in a simple $CP^1$
model with a potential term which determines an appropriate boundary condition
at infinity.

\section{Monopole-antimonopole solution in $CP^1$ model with a potential term}

We consider monopole-antimonopole field configuration determined
by winding numbers $m=1, n=2$ in a simple case when the function $f(r,\theta)$
entering the ansatz (\ref{ansmm}) vanishes identically.
Let us consider the following Lagrangian for the $CP^1$ model with
a potential term
\bea
{\cal L}&=& -\dfrac{1}{4}H_{\mu\nu}^2-V(u), \nn \\
V(u)&=& \dfrac{k^2}{4}\sin^2 \theta \cdot \nn \\
 && \Big (\dfrac{2}{2-i\sin \theta
(e^{-i \phi}u-e^{i\phi}\uast)}-\dfrac{1}{2}\Big )^2 .
\eea
Due to a special form of the chosen ansatz the potential term can be re-written in
a simple form in terms of the function $G(r,\theta)$
\bea
V(G)&=&\dfrac{k^2}{4}  \sin^2 \theta(G-\dfrac{1}{2})^2.
\eea
The potential term provides an appropriate boundary condition for the field $G(r,\theta)$
 at space infinity. A corresponding equation of motion
 represents a non-linear partial differential equation (pde)
\bea
&&\dfrac{1}{(1+\uast u)^2} \der_\nu (\sqrt g \der_\mu u H^{\mu\nu})-\dfrac{k^2}{2}r^2 \sin^3\theta
                                             (G-\dfrac{1}{2}) \nn \\
&&=0. \label{pde1}
\eea

The reduced form of the ansatz (\ref{ansmm}) with one non-vanishing function
$G$ implies the following expressions for the vector magnetic
field $\vec H_i=\frac{1}{2}\epsilon_{ijk}H_{jk}$ at space infinity
\bea
\vec H_r&=&-\dfrac{2 \sin (2 \theta)}{(1+\sin^2\theta)^2}, \nn \\
\vec H_\theta &=& 0,~~~~~ \vec H_\varphi=0.
\eea
The magnetic flux through the sphere $S^2$ of infinite radius
gives vanishing total magnetic charge
\bea
\int_{S^2} dr d\theta H_r=0.
\eea
The magnetic fluxes through the upper and lower half-spheres of $S^2$
are twice less compare to the monopole-antimonopole configuration
considered in the previous section
\bea
\Phi_+ &=& \int dr \int_0^{\frac{\pi}{2}} d\theta
       H_{r\theta}(\infty,\theta)=-2 \pi, \nn \\
\Phi_- &=& \int dr \int_{\frac{\pi}{2}}^\pi d\theta
       H_{r\theta}(\infty,\theta)=+2 \pi.
\eea
There is no magnetic flux around the $Z$-axis, consequently,
the Hopf charge density vanishes identically.

One can find solution near the origin $r\simeq 0$ and near space infinity
$r\simeq \infty$ using perturbation theory.
Expanding the function $G(r,\theta)$ in Taylor series
\bea
G(r,\theta)&=& g_1(\theta) r + g_2(\theta) r^2 + g_3(\theta) r^3 + ...,
\eea
one obtains a solution near $r=0$ with the following
coefficient functions up to third order of perturbation theory:
\bea
g_1(\theta)&=&c_1, \nn \\
g_2(\theta)&=&-(c_1^2+\dfrac{k^2}{256 c_1^2}), \nn \\
g_3(\theta)&=& \dfrac{7 c_1^3+3 c_2}{5}+\dfrac{9k^2}{640 c_1}(1-\dfrac{1}{2^{10} c_1^4}) \nn \\
             &+& c_2 \cos (2 \theta),
\eea
where $c_1,c_2$ are arbitrary integration constants.
Notice the appearance of the angle dependent term in the last equation which implies
that solution is axially symmetric and depends on two variables, $(r,\theta)$.
In asymptotic region near space infinity, $r\simeq \infty$, one has a
solution which is expressed by the series expansion (up to second order of perturbation theory)
\bea
&&G(r,\theta))=\dfrac{1}{2}+\sum_{n=1}^\infty b_n(\theta) \dfrac{1}{r^{4n}}, \nn \\
&&b_1(\theta)=\dfrac{2048(7+3\cos(2\theta))}{k^2(\cos(2\theta)-3)^5}, \nn \\
&&b_2(\theta)=\dfrac{134217728}{k^4(\cos(2 \theta)-3)^{11}} \nn \\
&&\cdot (119-235 \cos(2\theta)-139 \cos(4\theta)-9\cos(6\theta)).
\eea
The solution is given by asymptotic series and represents a non-perturbative solution
which exists only for non-zero values of the parameter $k$.
With this one can solve numerically the partial differential equation (\ref{pde1})
imposing Neumann boundary conditions $\pro_\theta G(r,\theta)|_{\theta=0,\pi}=0$.
The solution has been obtained by using the package COMSOL 3.5,
the plot for the function $G(r,\theta)$ is depicted in Fig. 7
with the model parameter $k=2$.
\begin{figure}[htp]
\centering
\includegraphics[width=80mm,height=60mm]{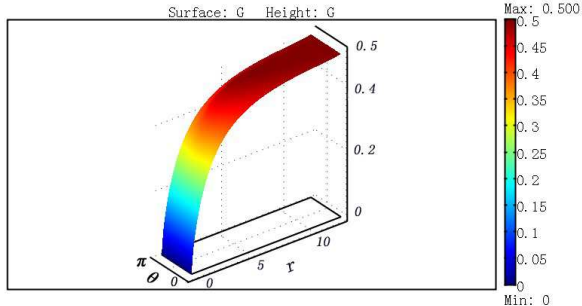}
\caption[plot1]{Solution for the function $G(r,\theta)$ with the model parameter $k=2$.}\label{fig2}
\end{figure}
\begin{figure}[htp]
\centering
\includegraphics[width=80mm,height=60mm]{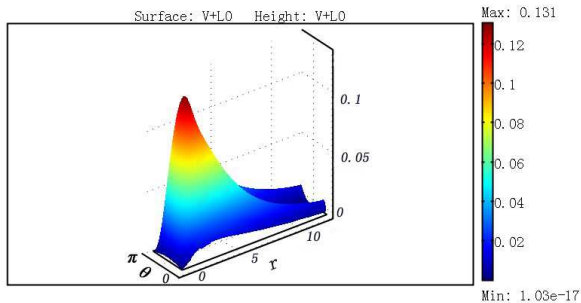}
\caption[plot1]{Total energy density plot, $k=2$.}\label{fig3}
\end{figure}
The energy density profile, Fig. 8, has one local maximum with two tails falling down
at far distance. The solution represents non-helical
magnetic field configuration, $\vec H_\varphi=0$.

One can solve PDE for various values of the model parameter $k$.
For $k \geq 0.5$ one has good convergency properties for all solutions.
The ratio of the potential energy $E_{pot}^{pde}$ to the total energy
$E_{tot}^{pde}$ for various values of the parameter $k$ is near $0.44$.
The obtained total energy values
are in good agreement with estimates obtained by using variational method
of minimizing the energy functional, see Table 1.
\vspace{5 mm}
\begin{table}
\begin{tabular}{ |l | c | c | c | }
   \hline
  ~ k & $E^{pde}_{total}$ & $E_{pot}^{pde}/E_{total}^{pde}$ & $E_{total}^{var}$  \\ \hline
  ~ 0.5 & 3.68  &0.45&3.42 ~\\ \hline
  ~ 1 & 5.31 &0.44 & 5.14 ~\\ \hline
  ~ 2 & 7.61 &0.44&7.92~ \\ \hline
  ~ 5 & 12.22 &0.44 &15.58~ \\ \hline
 \end{tabular}
\caption{Energy values for different values of the parameter $k$:
$E_{total}^{pde},~E_{pot}^{pde}$ are the total and potential energies
retrieved from solving pde, $E_{total}^{var}$ is the total energy
obtained by using variational method}
\end{table}

In conclusion, we have proposed a new type of finite energy
monopole configuration which can be treated as a monopole-antimonopole pair.
An essential feature of the configuration is that it does not
possess localized magnetic charges whereas the magnetic fluxes through
the upper (lower) half-spheres of infinite radius correspond
to monopole (antimonopole) charge. The discrete values of the magnetic
charges are conditioned by integer winding numbers $(m,n)$,
whereas a helical structure of the magnetic field is provided
by non-zero value of the Hopf charge.
Such finite energy monopole-antimonopole configurations minimize the energy functional
in restricted QCD, and they can play an important role in QCD.

The existence of monopole-antimonopole solution in a simple $CP^1$
model shows that the field $\hat n$ can regularize the singularity
inherent to point like monopoles. In standard QCD the field $\hat n$ represents
pure topological degrees of freedom, i.e., it does not have dynamic content.
Due to this, rather there is no monopole solution with non-zero magnetic charge in pure QCD.
However, in effective theories of QCD describing infrared limit
in Faddeev-Niemi formalism
\cite{FNprl99,shabanov} the field $\hat n$
manifests dynamical properties.
It would be interesting to study extended Skyrme-Faddeev-Niemi
models with potential terms \cite{adam,foster,ferreira}
in search of possible monopole-antimonopole solutions.
Another important implication of our results is related to the problem
of existence monopoles in the theory of electro-weak interactions.
We expect that finite energy monopole-antimonopole solution can
exist within the formalism of the standard model.
These issues will be considered in a subsequent paper \cite{pzzRC4}.

\acknowledgments
One of authors (DGP) thanks Prof. Y.M. Cho, E.N. Tsoy for useful discussions.
The work is supported by
NSFC (Grants 11035006 and 11175215), CAS (Contract No. 2011T1J31),
and by UzFFR (Grant F2-FA-F116).

\end{document}